\documentclass[journal,comsoc]{IEEEtran}
\usepackage[utf8]{inputenc}
\usepackage{color}
\usepackage{soul}
\usepackage{acronym}
\acrodef{sn}[SN]{satellite network}
\acrodef{an}[AN]{aerial network}
\acrodef{gn}[GN]{ground network}
\acrodef{rf}[RF]{radio frequency}
\acrodef{ris}[RIS]{reconfigurable intelligent surface}
\acrodef{vlc}[VLC]{visible light communications}
\acrodef{re}[RE]{reflective element}
\acrodef{geo}[GEO]{geostationary orbit}
\acrodef{meo}[MEO]{medium earth orbit}
\acrodef{leo}[LEO]{low earth orbit}
\acrodef{hap}[HAP]{high altitude platform}
\acrodef{lap}[LAP]{low altitude platform}
\acrodef{vn}[VN]{virtual network}
\acrodef{uav}[UAV]{unmanned aerial vehicle}
\acrodef{3d}[3D]{three-dimensional}
\acrodef{fso}[FSO]{free-space optical}
\acrodef{owc}[OWC]{optical Wireless Communications}
\acrodef{bs}[BS]{base station}
\acrodef{ber}[BER]{bit error rate}
\acrodef{thz}[THz]{terahertz}
\acrodef{mmwave}[mmWave]{millimeter wave}
\acrodef{ris}[RIS]{reconfigurable intelligent surface}
\acrodef{mimo}[MIMO]{multiple-input multiple-output}
\acrodef{los}[LoS]{line-of-sight}
\acrodef{nlos}[NLoS]{Non-LoS}
\acrodef{ir}[IR]{infrared}
\acrodef{uv}[UV]{ultraviolet}
\acrodef{ioe}[IoE]{Internet-of-Everything}
\acrodef{occ}[OCC]{optical camera communication}
\acrodef{lifi}[LiFi]{light fidelity}
\acrodef{led}[LED]{light-emitting diode}
\acrodef{ml}[ML]{machine learning}
\acrodef{qos}[QoS]{quality-of-service}
\acrodef{qoe}[QoE]{quality-of-experience}
\acrodef{sagin}[SAGIN]{space-air-ground integrated network}
\acrodef{csi}[CSI]{channel state information}
\acrodef{em}[EM]{electromagnetic}
\acrodef{gan}[GAN]{generative adversarial network}
\acrodef{dt}[DT]{Digital Twin}
\acrodef{snr}[SNR]{signal-to-noise ratio}
\acrodef{sinr}[SINR]{signal-to-interference-plus-noise ratio}
\acrodef{e2e}[E2E]{end-to-end}
\acrodef{5g}[5G]{fifth generation}
\acrodef{mec}[MEC]{mobile edge computing}
\acrodef{mems}[MEMS]{microelectromechanical systems}
\acrodef{6g}[6G]{sixth generation}
\acrodef{b5g}[B5G]{beyond 5G}
\acrodef{tbps}[Tbps]{Tera-bit-per-second}
\acrodef{ct}[CT]{cyber twin}
\acrodef{pt}[PT]{physical twin}
\acrodef{ai}[AI]{artificial intelligence}
\acrodef{urllc}[URLLC]{ultra-reliable low latency communication}
\acrodef{iiot}[IIoT]{industrial IoT}
\acrodef{isac}[ISAC]{integrated sensing and communication}
\acrodef{iot}[IoT]{internet-of-things}
\acrodef{kpi}[KPI]{key performance indicator}
\acrodef{fov}[FoV]{field of view}
\acrodef{nfv}[NFV]{network function virtualization}
\acrodef{sdn}[SDN]{software defined networking}
\acrodef{mura}[MURA]{massive unsourced random access}
\acrodef{hetnets}[HetNets]{heterogeneous networks}
\acrodef{gnn}[GNN]{graph neural networks}
\acrodef{jscc}[JSCC]{joint source and channel coding}
\acrodef{ofdm}[OFDM]{orthogonal frequency division multiplexing}
\acrodef{m2m}[M2M]{machine-to-machine}
\acrodef{oam}[OAM]{orbital angular momentum}

\usepackage[justification=centering]{caption}

\usepackage[pdftex]{graphicx}

\title{Digital Twin-Empowered Communications: A New Frontier of Wireless Networks}

\begin{document}
\author{Lina~Bariah, Hikmet~Sari, and M{\'e}rouane~Debbah

%\vspace{-0.9cm}
}

\maketitle
 
\begin{abstract}
The future of wireless network generations is revolving toward unlocking the opportunities offered by virtualization and digitization services, with the aim to realize improved quality-of-experience (QoE) and bring several advantages to network users. According to the rapid development in the field of network virtualization, we envision that future wireless networks will run over ubiquitous deployment of virtualized components that are controlled by artificial intelligence (AI), i.e., the conceptualization of the \textit{Digital Twin} (DT) paradigm. The key principle of the DT relies on creating a holistic representation of wireless network elements, in addition to decoupling the information pertaining to physical objects and dynamics, into a cyber twin. The cyber twin will then leverage this information for AI models training, and then reasoning and decision-making operations, which will be then reflected to the physical environment, for improved sustainability. Motivated by this, in this article, we dig deep into the intertwined role of wireless technologies as being enablers and enabled by the DT. Furthermore, we put a forward-looking vision of the integral role that future 6G networks are anticipated to play in order to realize an efficient DT. Finally, we sketch the roadmap toward identifying the limitations of the DT in 6G-enabled wireless networks, and open new horizons for further developments in different design aspects. 
\end{abstract}

\section{Introduction} 

Since the evolution of smart city concept, the way we are living our everyday life has been revolving toward integrating intelligence in every aspect of our lives.
%, spanning from automated systems, smart homes, smart buildings and factories, to intelligent transportation systems and intelligent retail. 
While this paradigm has motivated the emergence of novel technological trends, in order to meet its demanding requirements, several technologies have played a fundamental role in enabling smart city concept, such as \ac{iot}, sensing, \ac{ml} and \ac{ai}, and cloud and edge computing. Along the way, wireless network generations have fundamentally participated in maturing the paradigm of smart city, which will continue to grow with the aim to realize pervasive user-centric applications supported by ubiquitous intelligence. This means, \ac{ai} is anticipated to be the fuel for every aspect in smart cities. Hence, it is necessary to further explore what beyond 5G networks will bring, in terms of new technologies and services, in order to deliver the promised \ac{qoe} to network users.  
%\begin{table*}
%\centering
%\caption{}
%\resizebox{0.7\linewidth}{!}{\includegraphics{Literature.png}}
%\label{hard}
%\end{table*}
While the research continues to argue what future wireless generations will be, it has become apparent that 6G will be shaped towards provisioning network virtualization and softwarization, with the aim to support ubiquitous deployment of latency-sensitive applications \cite{bariah2020prospective}. While ultra-high reliability and data rates are essential for the successful implementation of these applications, they require the development of online proactive mechanisms, in order to realize self-adaptive, self-optimizing, and self-sustaining networks, with intelligent decision-making capabilities. In this regard, \ac{dt} has recently been identified as a promising candidate for enabling zero-touch wireless networks, thanks to \ac{sdn} and \ac{nfv} which have paved the way for the evolution of interactive physical-cyber platforms. The key principle of the \ac{dt} paradigm is to create a virtual representation for the physical elements and the dynamics and functions of the network. According to its definition, the \ac{dt} is envisioned to enable end-to-end digitization of wireless networks, with the aim to perform cost-effective, adaptive, and fast network-wide optimization and design \cite{wu2021digital}. Furthermore, the \ac{dt} allows the utilization of the digital realm with the aim to develop and test novel schemes and \ac{ai} algorithms, that are capable of handling previously experienced or envisioned scenarios based on the collected data at the \ac{ct}, and then to implement them at the \ac{pt} once fully mature. 

Despite its promising advantages, to reap the full potential of the \ac{dt} technology in 6G networks, the \ac{ct} is envisaged to leverage \ac{ai} algorithms, with novel data-driven paradigms, high performance computing, optimization theory, matching theory, and efficient cyber-physical interaction schemes, to realize the necessary adaptation/reconfiguration at the \ac{pt} with an imperceptible time-lag. For the successful implementation of a high-fidelity \ac{dt} paradigm in 6G, a new level of stringent requirements pertaining to connectivity, reliability, latency, and data rate are imposed on future wireless generations. In particular, it is foreseen that \ac{dt} will require an outage time/year less than 52 minutes, with 99.99\% reliability. Furthermore, sufficient bandwidth is necessary to handle the data transfer requirements between the \ac{pt} and \ac{ct}, especially for real-time applications that require low-latency communication. Also, collected data from the twin necessitates the development of security schemes for ensuring trusted data repositories, as well as hardware \& software security of data collectors. Moreover, \acp{dt} often involve the integration of various systems and devices, which may operate on different network protocols or communication standards. The network infrastructure should support interoperability and provide the necessary protocols and interfaces to enable seamless connectivity and data exchange between different components of the \ac{dt} ecosystem. Despite being in its infancy, few standardization and industrial activities on \ac{dt} have been released by different entities (Table \ref{tab:stand}).

%\begin{table}
%\caption{DT Standardization Activities} \label{tab:stand} 
%\begin{tabular}{ | c | c | p{15em} | } 
%  \hline
%  Entity/Number & Date & Scope \\
%  \hline 
%   ITU-T Y.3090 & (02/2022) & \begin{itemize}
%       \item Description of network DT's architectures
%       \item Functional \& Service requirements of network DT.
%       \item Security consideration of network DT.
%   \end{itemize} \\ 
%  \hline
%   IEFT & (03/2023) & \begin{itemize}
%       \item DT definition
%       \item Benefits of DT.
%       \item Challenges of network DT.
%       \item Enabling technologies to build DT network.
%       \item Applications and security consideration.
%       \item DT interaction with intent-based networking.
%       \end{itemize}\\ 
%  \hline
%  3GPP & (08,11/2021) & \begin{itemize}
%      \item Scenarios, corresponding requirements and possible solutions when integrating DT into network management system (under development)
%  \end{itemize}	\\
%  \hline
%  \hline
%\end{tabular}
%\end{table}

\begin{table}
\caption{DT Standardization Activities} \label{tab:stand} 
\begin{tabular}{ | c | c | p{15em} | } 
  \hline
  Entity/Number & Date & Scope \\
  \hline 
   ITU-T Y.3090 & (02/2022) & \begin{itemize}
       \item Description of network DT's architectures
       \item Functional \& Service requirements of network DT.
       \item Security consideration of network DT.
   \end{itemize} \\ 
  \hline
   IEFT & (03/2023) & \begin{itemize}
       \item DT definition
       \item Benefits of DT.
       \item Challenges of network DT.
       \item Enabling technologies to build DT network.
       \item Applications and security consideration.
       \item DT interaction with intent-based networking.
       \end{itemize}\\ 
  \hline
  3GPP & (08,11/2021) & \begin{itemize}
      \item Scenarios, corresponding requirements and possible solutions when integrating DT into network management system (under development)
  \end{itemize}	\\
  \hline
  \hline
\end{tabular}
\end{table}

%Accordingly, the research on the advancements of \ac{dt} and its integration into wireless networks have picked up the pace, where several contributions in the literature have tackled the implementation of \ac{dt} in wireless networks, from different perspectives. %The article outline is demonstrated in Fig. \ref{fig:outline}.

\begin{table}
\caption{DT Wireless Networks: State-of-the-Art} \label{tab:soa} 
\begin{tabular}{ | c | p{25em} |  } 
  \hline
  Reference & Topics covered\\
  \hline 
  \cite{wu2021digital} & \begin{itemize}
      \item DT definition.
      \item Cyber-physical communication requirements.
      \item Cloud and edge computing.
      \item DT applications, including, aviation, manufacturing, 6G, healthcare, etc.
  \end{itemize}  \\ 
  \hline
  \cite{nguyen2021digital} & \begin{itemize}
      \item AI for DT. 
      \item DT for 5G automotive, 5G radio and channel emulation, and 5G optimization and testing.
  \end{itemize}  \\ 
  \hline
  \cite{tang2022survey} & \begin{itemize}
      \item Basic concept of \ac{dt} and \ac{mec}.
      \item Advantages of \ac{dt} edge networks: communication and computation perspectives. 
      \item Cloud, edge, and hybrid deployment of \ac{dt}.
      \item \Ac{ml} and blockchain in \ac{dt} edge networks.
  \end{itemize}  \\ 
  \hline
  \cite{luan2021paradigm} & \begin{itemize}
      \item \Ac{dt} structure.
      \item Edge and cloud deployment of \ac{dt}.
      \item Edge- and cloud-computing based \ac{dt}.
  \end{itemize} \\ 
  \hline
  \cite{khan2022digital} & \begin{itemize}
      \item Key design requirements of \ac{dt}.
      \item Architecture of \ac{dt}-enabled 6G.
      \item Operational steps of \ac{dt}-enabled 6G.
  \end{itemize}  \\ 
  \hline
  \cite{khan2022digital2} & \begin{itemize}
      \item Key concepts of \ac{dt}.
      \item Taxonomy of \ac{dt} for wireless networks: twins isolation, decoupling, interfaces design, twin objects design, etc.
      \item The role of wireless networks in \ac{dt}: air interface, edge devices, and security \& privacy.
  \end{itemize}  \\ 
  \hline 
  \cite{lin20226g} & \begin{itemize}
      \item \ac{dt} use-cases in wireless networks, e.g., network simulation \& planning, network operation \& management, what-if 6G analysis, etc.
      \item 6G \ac{dt} network requirements.
      \item \ac{dt} network architecture.
  \end{itemize}  \\ 
  \hline 
  \cite{ahmadi2021networked} & \begin{itemize}
      \item Key design elements of \ac{dt}.
      \item Advantages of \ac{dt} in wireless networks.
      \item Standardization of \ac{dt}.
      \item 6G requirements to enable \ac{dt}.
  \end{itemize}  \\ 
  \hline
  \hline
\end{tabular}
\end{table}

\begin{figure*}
	\centering
	\resizebox{1\linewidth}{!}{\includegraphics{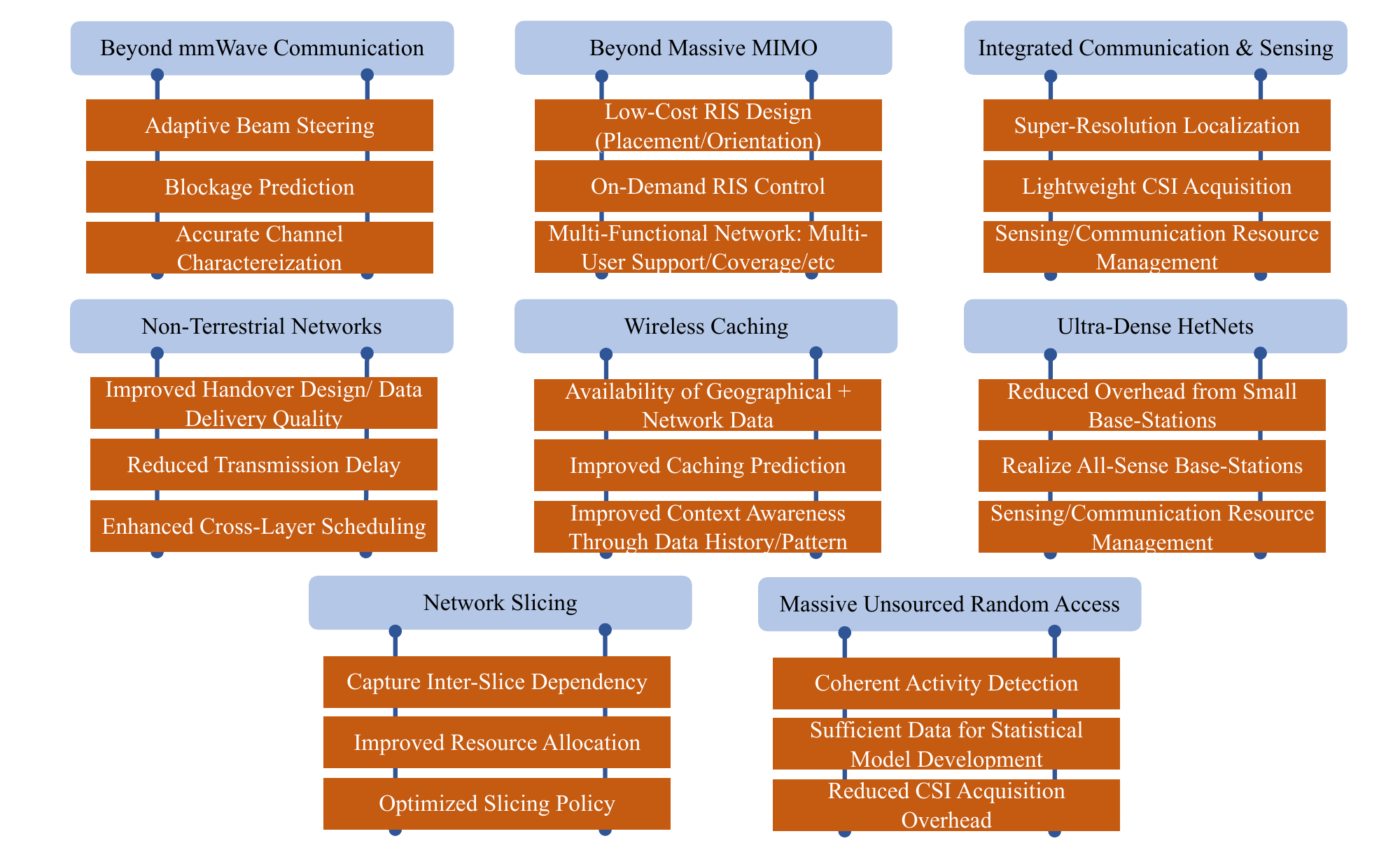}}
	  \caption{The role of DT in wireless networks.}
    \label{fig:DT4WLS}
\end{figure*}

\subsection{Related Work}

Although several research contributions have discussed the implementation of \ac{dt} within the context of wireless networks, e.g. \cite{wu2021digital}-\cite{fuller2020digital}, in this article, we aim to shed light on how to unlock the full potential of different wireless technologies, through leveraging the \ac{dt}. In particular, in \cite{wu2021digital}, the authors defined the \ac{dt} paradigm, discussed the key requirements of physical-virtual communications, and identified the various requirements of \ac{pt}-\ac{pt}, \ac{pt}-\ac{ct}, and \ac{ct}-\ac{ct} links. The article \cite{nguyen2021digital} focuses mainly on how \ac{ai} will introduce a noticeable enhancement to the performance of \ac{dt} in 5G networks. Particularly, the authors have focused on three main concepts, 5G automotive, 5G radio and channel emulation, and 5G optimization and validation, and how edge intelligence can support such concepts. In \cite{tang2022survey}, Tang \textit{et al.} articulated the basic concepts of \ac{dt} and \ac{mec}, and provided a general overview on the communication and computation advantages reaped from \ac{dt} edge network. They further investigated the technical deployments of \ac{dt} at the cloud and at the edge. They also discussed blockchain and \ac{ml} paradigms within the \ac{dt} networks. The authors in \cite{luan2021paradigm} explored the \ac{dt} structure, and discussed the edge and cloud deployment of \ac{dt}. The key design requirements were discussed in detail in \cite{khan2022digital}, where the authors studied the decoupling, intelligent analytics, blockchain-based data management, and scalability and reliability. Also, they presented a general architecture of the \ac{dt} when implemented in 6G, and highlighted the main operation steps. On the other hand, the authors in \cite{khan2022digital2} articulated the key concepts of \ac{dt}, and provided a taxonomy for the implementation of \ac{dt} in wireless networks. Also, they discussed the role of wireless network in realizing the vision of \ac{dt}. Finally, the authors in \cite{fuller2020digital} have provided a comprehensive survey on the definition of the \ac{dt}, and its applications (e.g., smart city, manufacturing, and healthcare). They also shed light on the utilization of \ac{dt} in industry, where they highlighted the challenges pertaining to data analytics, \ac{iot}, and \ac{dt} design. Additionally, they articulated the history behind \ac{iot} networks, \ac{ml}, and the \ac{dt}. The state-of-the-art is summarized in Table \ref{tab:soa}. 

\begin{figure}
	\centering
	\resizebox{1\linewidth}{!}{\includegraphics{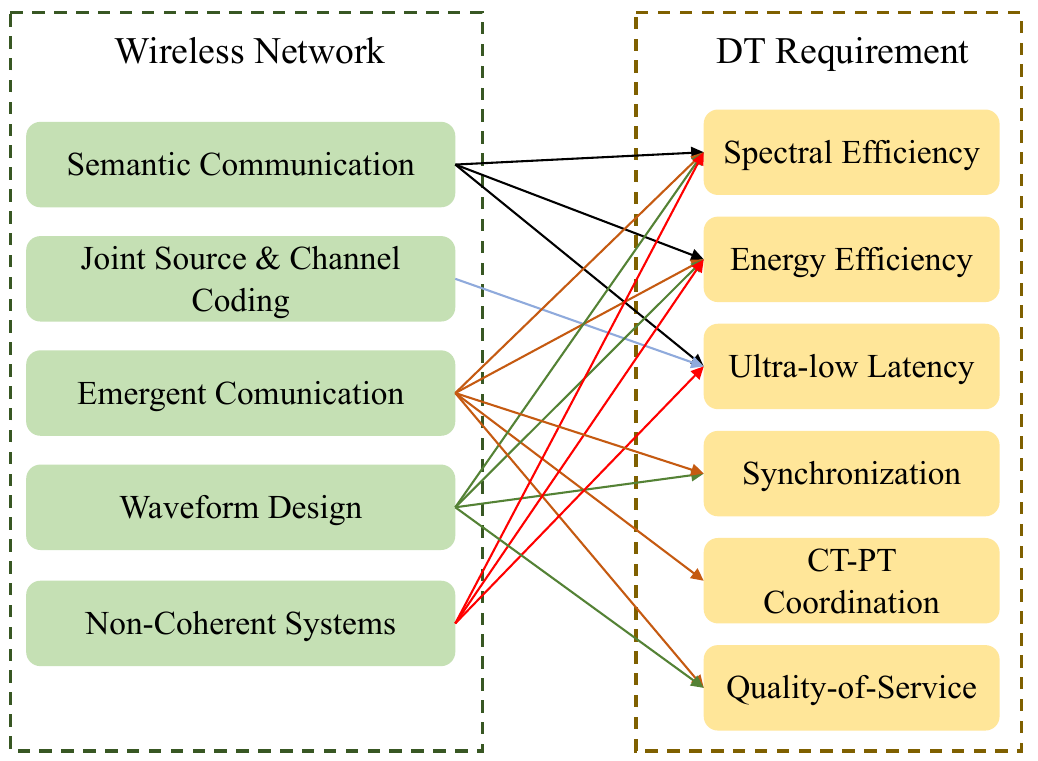}}
	  \caption{The role of wireless networks in DT.}
    \label{fig:WLS4DT}
\end{figure}

Different than the existing literature, the objective of this article is three-folds. First, we overview the technological trends that have motivated the concept of \ac{dt}-empowered 6G, and have defined the constraints and requirements that need to be taken into consideration when designing \ac{dt} networks. Second, we identify the encountered challenges in different wireless technologies, and we
put a vision forward on how to employ the \ac{dt} paradigm in order to overcome such challenges and introduce several advantages to different wireless technologies. Third, we approach the interplay of \ac{dt} and 6G from fundamental limitations point of view, where we explore the theoretical limitation of semantic communication, air-interface design, source and channel coding design, etc., on the implementation of \ac{dt}. We further open the floor for exploring \ac{dt} as an enabler for 6G through stressing on some challenges that are envisioned to be encountered when such a technology is deployed. Note that such a discussion has not been presented the current literature. In Figs. \ref{fig:DT4WLS} \& \ref{fig:WLS4DT}, we summarize the intertwined role of \ac{dt} and wireless networks in providing enhanced user's experience.

%\begin{figure*}
%	\centering
%	\resizebox{0.8\linewidth}{!}{\includegraphics{Literature.png}}
%	  \caption{Summary of existing related work.}
%    \label{fig:lit}
%\end{figure*}

%\section{\color{blue}Overview of Digital Twin}

%\section{\color{blue}Enabling Technologies of Digital Twin}
%\begin{enumerate}
%    \item \textcolor{blue}{Network Slicing}
%    \item \textcolor{blue}{Software-Defined Networks (SDNs)}
%    \item \textcolor{blue}{Network Function Virtualization (NFV)}
%    \item \textcolor{blue}{Machine Learning}
%    \item \textcolor{blue}{Cloud Computing}
%\end{enumerate}
\section{Digital Twin-Empowered Wireless Networks: Driving Trends}

\subsection{Sustainable Smart Cities}

New smart cities are evolving toward conceptualizing urban sustainability principle, by fueling the initiatives of building green cities, that are envisioned to enjoy zero harmful emissions and reduced carbon footprint, and improved resources utilization and waste management efficiency. While wireless networks in smart cities aim at achieving an improved life quality, and to deliver the needed \ac{qoe} for city users, it is essential to revisit current wireless technologies in order to develop environment-friendly system that support sustainability in smart cities. A promising candidate, the \ac{dt} technology promises to enable sustainable cities, through performing a virtualized network planning, optimization, configuration, and adaptation, and hence, extremely reducing the consumption of the network resources. Route selection in autonomous driving constitutes a potential use-case for sustainable \ac{dt}-empowered cities, where a self-optimized route selection can help, not only in delay reduction and congestion avoidance, but in an improved efficiency in the vehicles' resources as well. Nevertheless, due to the heterogeneous nature of smart city applications, from healthcare and intelligent manufacturing to urban planning and congestion control, large data sets arises from the continuous monitoring of the network that need to be processed, and then, heavy computing tasks need to be executed for performing multi-objective optimization of the network resources, for guaranteed reliability and ensured optimum solutions. 
%\begin{figure}
%	\centering
%	\resizebox{0.8\linewidth}{!}{\includegraphics{outline.pdf}}
%	  \caption{\textcolor{blue}{The outline of the article.}}
%    \label{fig:outline}
%\end{figure}

\subsection{Zero-touch Networks}
The fourth industrial revolution, Industry 4.0, has conceptualized the vision of automation in \ac{iiot}, and has motivated the emergence of zero-touch network management, and hence, the unfolding of self-healing networks that are capable of autonomously identifying experienced faults and perform real-time self-maintenance, without human intervention. With the diverse fault scenarios in smart cities, including electricity supply interruption, unplanned massive traffic congestion, and unpredicted security breach, current fault management techniques might be insufficient to handle the increased operational complexity, and to realize closed-loop automation for network management. In this regard, the \ac{dt} technology can provide the necessary flexibility feature to enable automated fault management and alleviate the management overhead from the network nodes. By leveraging \ac{ai} algorithms, multiple agents can be trained in the \ac{ct} over a very wide range of possible network failures and faults, in order to be ready to overcome potential incidents, and to autonomously perform qualitative faults detection, isolation, and restoration, when installed in the physical environment. Besides reduced physical overhead, other promising advantages can be revealed when \ac{dt} is employed for enabling self-healing networks. Generally speaking, despite their severe effect on the network status, disasters and critical network faults rarely happen in real cities, and therefore, the available data for agents training is limited. Therefore, the \ac{ct} can be exploited to generate synthetic data that represent a diverse range of fault scenarios, and therefore, enable improved agents/models training. 

\subsection{URLLC}

With the strong believe that the role of virtual and augmented reality services is instrumental in achieving the \ac{qos} requirements of emerging applications, \ac{urllc} has been identified as one of the 6G verticals, and an enabler for immersive real-time metaverse. Specifically, in order to realize the envisioned seamless \ac{dt} paradigm, sensing, communications, and computing tasks are anticipated to be performed in few milliseconds, and the \ac{ct} is expected to be in a perfect synchronization with its physical counterpart. Equally important, extremely reliable communication should be guaranteed in \ac{dt}-empowered wireless networks, in order to ensure accurate model training, and hence, precise decision-making process. Within the context of smart cities, \ac{urllc} is an essential element in deploying mission-critical applications, such as autonomous driving, remote surgery, and public safety, to name a few \cite{van2022edge}. \ac{dt} and \ac{urllc} technologies in smart city applications are intertwined, in the sense that each is an enabler and being enabled by the other. Taking the example of autonomous driving, on one hand, instead of imposing a huge pressure on the network energy and spectrum resources, and draining the limited on-board resources of smart vehicles and roadside units to perform intelligent prediction and fast inference, the parallel cyber world can be leveraged in order to perform lightweight control over autonomous vehicles. Furthermore, parallel driving can be an efficient approach to overcome the short-sight challenge in vehicular networks, where vehicles have limited knowledge of the network status, constrained by the limited coverage of vehicles, and hence, the \ac{dt} can provide a holistic representation of the whole monitored area, over a long period of time. This enables the network controller to develop a sufficient level of understanding of the network dynamics and scenarios, and therefore, perform reliable and fast control over future actions. On the other hand, the successful implementation of the \ac{dt} paradigm in autonomous driving scenarios is constrained by the efficiency of the cyber-physical interface, and its ability to deliver the needed on-demand sensing, modeling, and communications, within a constrained time-frame and with the required accuracy, in order to realize safe autonomous driving.       
\begin{figure}
	\centering
	\resizebox{1\linewidth}{!}{\includegraphics{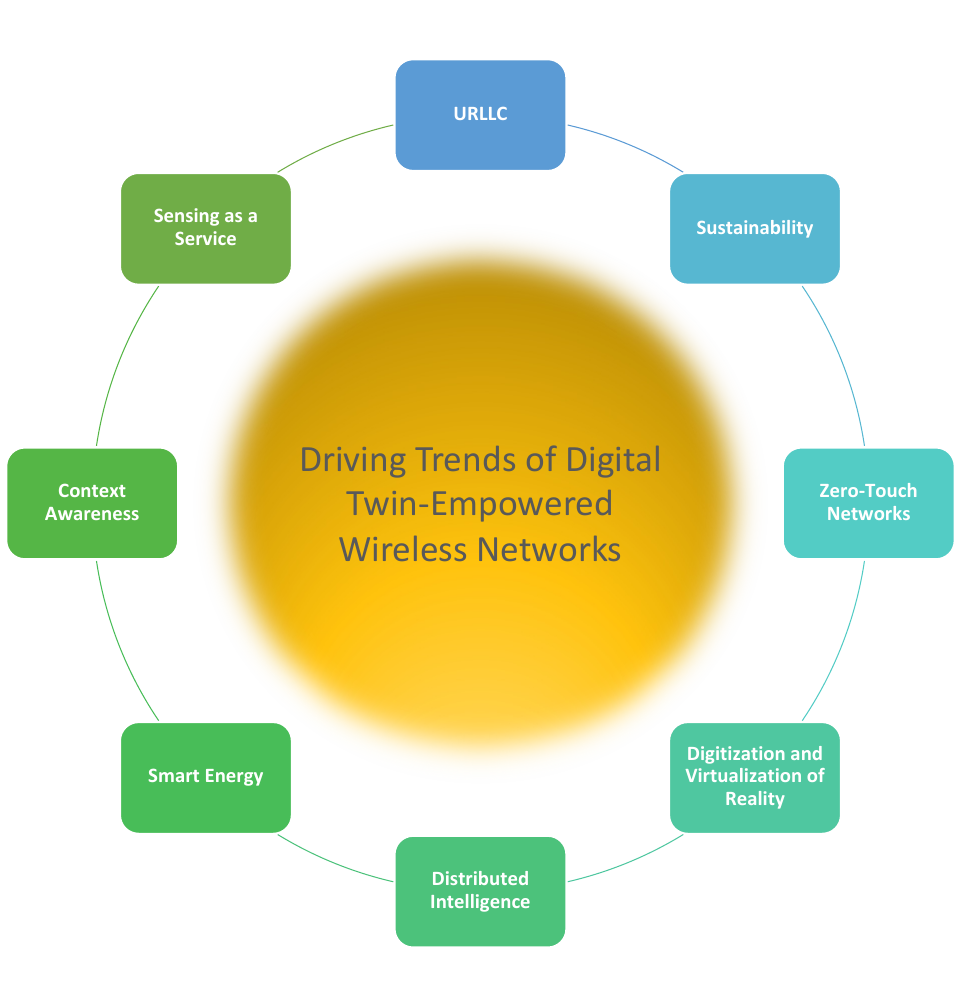}}
	  \caption{Key technological trends of DT-empowered wireless network.}
    \label{fig:trend}
\end{figure}

\subsection{Digital Representation of Reality}
Managing wireless networks is inherently complicated, with the extremely large number of homogeneous interactive nodes, and the existence of static and moving smart objects. Therefore, it is very challenging to enable network-wide control and to realize the envisioned automation in future smart cities. With the recent advancements in virtual and augmented reality services, the convergence of the physical and digital worlds has become feasible. While 5G has set the scene towards ultimately enabling cyber-physical interactions, 6G networks are envisioned to witness an evolution of network digitization paradigms, where digital representations of an entire networked city can be realized, incorporating each sophisticated aspect in highly dynamic cities. The utmost goal of such a digital platform is to enable a precise planning of the network assets and the execution of the necessary actions, with a distributed control capabilities \cite{Erc-DT}. While the digital representation should be in perfect synchronization with the physical realm, it gives the freedom to travel in time to the past or the future, with the aim to facilitate better understanding for the network dynamics, and predict future events for enhanced network planning.

In more details, a wireless \ac{dt} can enable the exploration of past and future scenarios to allow improved network planning, design, and configuration, as well as the development of tuned \ac{ai} models that can accommodate these scenarios. From the one hand, \acp{dt} can be used as a tool to acquire historical data, which can be leveraged to analyze past network performance, patterns, and events. Gaining insights into how the network has behaved in the past, helps identify trends, patterns, and potential issues that occurred, enabling better understanding of the network dynamics and behavior. On the other hand, an efficient \ac{dt} can be used to create rare network scenarios that might be experienced in future, allowing minimized risk and network failures. In particular, the exploitation of historical data from the \ac{ct} and real-time data from the \ac{pt}, can be integrated with simulated scenarios at the \ac{ct} to enable predictive analysis, which helps to identify potential bottlenecks, and predict events that may impact network operations. Such type of \acp{dt} are further aimed for generating data pertinent to future envisioned events, and therefore, they allow the realization of more generalized \ac{ai} models that are capable of handling a wide-range of network scenarios.

\subsection{Sensing as a Service}

Although sensing has already become an indispensable tool in current and future wireless networks, the \ac{dt} technology is envisaged to substantially participate in advancing the sensing capabilities, particularly in smart cities applications. This is motivated by the fact that sensors in \ac{dt}-enabled smart cities are required to perform a wide range of sensing functionalities in order to accurately and continuously capture all operational and environmental information that assist with understanding the dynamics of the network, and thereby allow the \ac{ct} to realize efficient data analysis and decision making. Therefore, it is essential for the sensory data that constitutes the fuel for the \ac{dt} paradigm to comprise diverse types of data, including vision-based data, voice-based data, control information, environmental status, activities and objects tracking, etc. As a result, current sensing tools, e.g., cameras, microphones, and inertial measurements tools, need to be redesigned as their probability of failure is expected to be high when integrated in \ac{dt}-empowered smart cities applications. This is due to several limitations, including their constrained battery-lifetime and high battery recharging/replacement overhead, as well as the strict sensed data-type and sensing frequency. Such limitations prevent delivering on-demand sensing, yielding inaccurate representation of the \ac{pt}. Accordingly, developing efficient, sustainable, and self-powered sensors that enjoy various operational modes, different material and compatibility features is essential for the successful deployment of \ac{dt}. While wearables and ultra-small sensors are a perfect fit for current sensing services, the emergence of \ac{dt}-empowered cities opens new horizons for more innovative solutions of efficient sensors that are embedded, implanted, or printed, in order to serve a diversity of applications with a high number of sensing modalities \cite{zhang2022artificial}.  

\subsection{Distributed Intelligence}

While building a holistic digital representation of a city is the vision, it might be challenging to achieve the needed \ac{qoe} through a single twin, due to the highly dynamic and heterogeneous nature of urban cities. This is particularly pronounced in highly-densed sprawling cities, where models training and inference experience long delays. By leveraging distributed learning algorithms, multiple distributed twins can be trained over multiple locations, and then a centralized twin can be developed for models aggregation and global decision making. Although such a distributed approach can help with overcoming the slow training over extremely-large data sets, concerns pertaining to multiple-twins coordination and accuracy of the trained model might arise, which might be intolerable by accuracy-sensitive life-threatening applications, e.g., autonomous driving. Furthermore, this requires the development of efficient twin-to-twin interfaces to enable successful interaction between the distributed multiple twin. Accordingly, a delay-accuracy trade-off problem is resulted.  

On the other hand, as distributed learning is an enabler for \ac{dt}, the latter can be leveraged with the aim to achieve lightweight distributed learning in vehicular edge networks. In particular, the \ac{ct} can be employed in order to train multiple agents, by utilizing the data collected from the network, and then the agents can be installed in the \ac{pt} for on-demand inference. In this approach, the agents are trained offline to perform the assigned tasks, ensuring reduced complexity and energy consumption at edge vehicles. Although the agents might need to interact with the physical world few times, the ultimate goal is to build a comprehensive \ac{dt} framework, which is capable of offering extensive model training that encompasses a wide-range of networks scenarios, and hence, an interaction with the physical environment will not be required.  

\subsection{Smart Energy}

With the evolution of the \ac{ioe} paradigm, and the emergence of human-centric urban and rural applications, smart energy management has been deemed as an instrumental key to allow wireless networks to enjoy a long life-cycle, contributing to the conceptualization of the sustainable city paradigm. Taking advantage of the surrounding environment, zero-energy devices unlock new opportunities in terms of energy efficiency and sustainability, where they are characterized by their no-battery architecture \cite{eric_ZE}. Rather, they exploit the vibrations, light sources, temperature gradients, and \ac{rf} signals to harvest the energy needed for signals communication, alleviating the overhead of battery replacement and recharging, particularly, in remote or hard-to-reach areas. The development of the zero-energy devices concept comes inline with the emergence of the \ac{dt} paradigm, where the former paves the way for the latter. From the \ac{dt} perspective, zero-energy devices represent the seed for on-demand, sustainable sensing services, in order to ensure an up-to-date synced \ac{ct}. Owing to their low-cost and low energy consumption, ubiquitous deployment of zero-energy devices can be implemented to allow continuous sensing of the \ac{pt} status, and hence, contribute to the successful realization of an operational \ac{dt}.%, which heavily relies on the timely sensing of the current network status, for accurate, on-demand inference and control. 

\subsection{Context Awareness}

Future wireless networks are envisioned to leverage context awareness for a better understanding of the environmental, temporal, and situational aspects of events and network behaviors. By offering intelligent storage and processing services, \ac{dt} introduces a new era of context awareness, where the fusion of multi-sensory data can be realized, and leveraged for recognizing nodes status and network traffic, and therefore, enables trends and deviation prediction. This can be achieved by understanding and storing the current and historical contexts, allowing real-time detection and identification of temporal deviations in nodes status. In this regard, wireless \ac{dt} offers an augmented dimension of data modality by integrating various sources of information and providing real-time insights. In particular, the data collected by \acp{dt} provides real-time insights into the behavior and performance of their physical counterparts, enabling proactive decision-making and anomalies detection, and thereby, improving overall context awareness. It is worthy to note that this data can include various types of information, such as operational parameters, environmental conditions, performance metrics, and user interactions. It should be further emphasized that the acquired data comprises images, videos, radio signals, and sound signals, offering an efficient platform for uncovering hidden patterns and correlations between the different data modalities, and therefore, contributing to a better understanding of the network behaviors. On the other hand, \ac{dt} constitutes a secure channel for context exchange, where sensitive data pertaining to nodes' status and location can be securely stored and processed at the twin.

\begin{figure*}
	\centering
	\resizebox{0.7\linewidth}{!}{\includegraphics{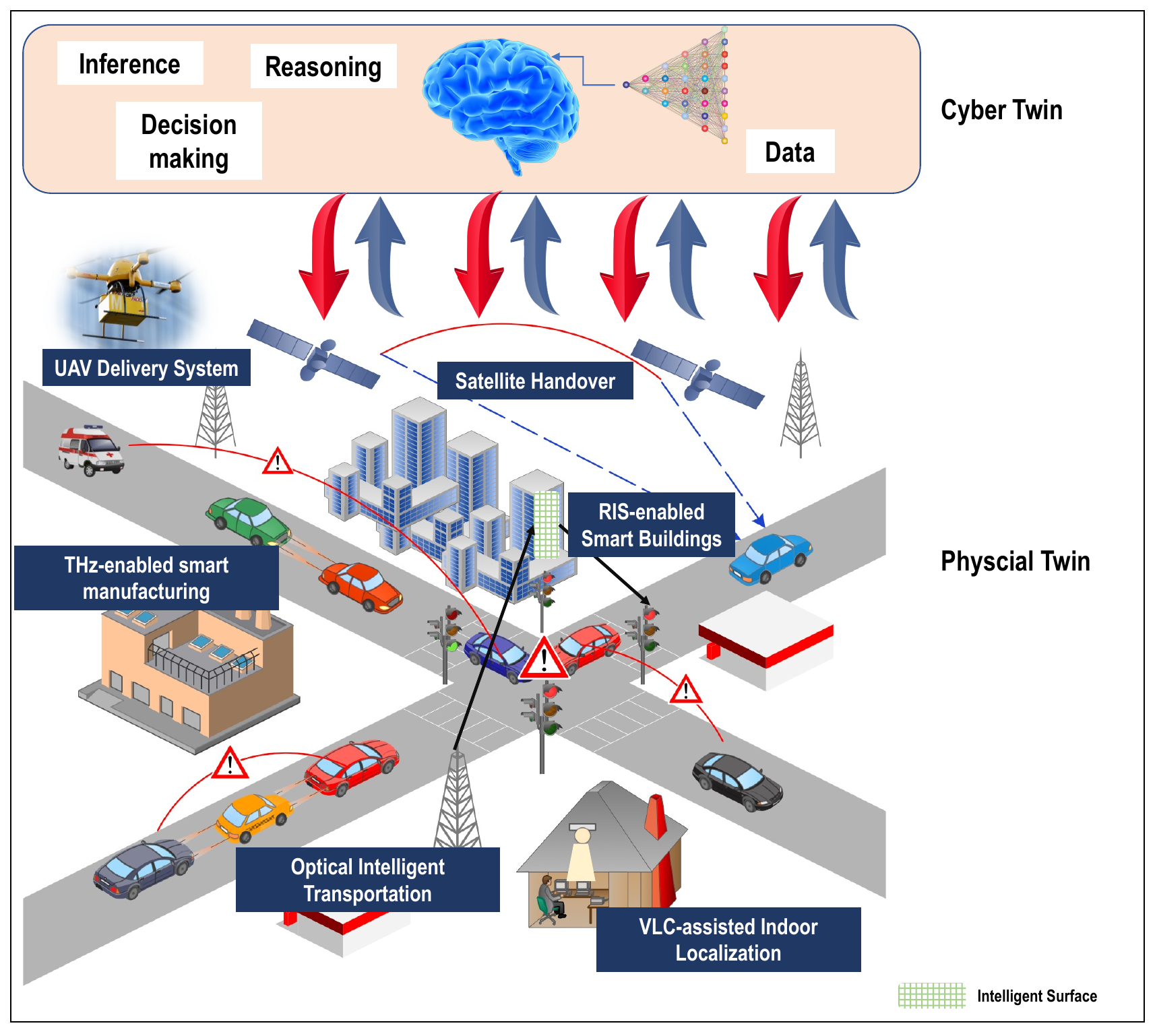}}
	  \caption{DT-enabled smart city: Wireless technology use-cases.}
    \label{fig:UC}
\end{figure*}
%\vspace{-5pt}
\section{Wireless Communication Technologies: How Digital Twin Fits}

\subsection{Beyond mmWave Communication}

While high frequency communication paradigms promise to offer several advantages in terms of latency and data rates, compared to their low frequency counterparts, several factors limit their applicability to key scenarios in future wireless generations, including the limited communication distances due to the high path-loss, the strong channel sparsity which limits the spatial multiplexing gains, and the dominance of the line-of-sight (LoS) component which makes high-frequency communication paradigms prone to signal blockage. In the following, we identify the potentials of \ac{dt} as an enabler for high-frequency communications, including, \ac{thz}, \ac{vlc}, and \ac{owc}.

\subsubsection{Terahertz Communications}

Albeit the promising potentials of \ac{thz} communications in future wireless generation, in terms of spectral and energy efficiency, and the wide range of 6G applications, that are anticipated to be supported by \ac{thz} links, e.g., indoor localization, vehicle-to-vehicle communication, smart manufacturing, and sensing, several concerns pertaining to the design and implementation of \ac{thz} communications need to be tackled. This includes the vulnerability of \ac{thz} links to deep fading and blockage, which results in unreliable \ac{thz} links and a degraded \ac{qos} for end users. This challenge exacerbates with the ultra-narrow beamwidth of \ac{thz} links, rendering \ac{thz} systems vulnerable to pointing errors, blockage, with limited coverage and multi-user support. It was recently demonstrated that \ac{dt} can be instrumental in enhancing the performance of \ac{thz} communication in indoor environments, even with highly dynamic networks, through enabling improved context awareness and efficient, adaptive beam steering \cite{pengnoo2020digital}. This is achieved through leveraging the \ac{dt} to model, predict, and control the \ac{thz} channel characteristics, in order to find the optimum signal path, and therefore, to maximize the received \ac{snr} at the receiver end. It should be highlighted that an accurate channel modeling at the \ac{thz} frequency is challenging, and despite the fact that there are several research activities initiated to derive accurate channel models that are experimentally tested as well as analytically modeled, this challenge is still an open research topic. Accordingly, the \ac{dt} offers a rich environment for testing and measuring \ac{thz} channels under various scenarios through heuristic approaches, and validating such models with theoretically derived models, for enhanced modeling accuracy.

\subsubsection{Visible Light Localization and positioning}

Over the last couple of years, extensive research and industrial efforts were devoted to exploit \ac{vlc} for localization and positioning purposes, due to its inherent low deployment cost, high throughput, and security features. While current visible light localization has demonstrated an acceptable performance in terms of 2D and 3D localization, it suffers from some challenges that prevent them from delivering the needed \ac{qos} imposed by future wireless generations. In addition to the ambient light noise, these challenges include the performance-degrading interference experienced when an array of \acp{led} is implemented. Furthermore, the limited flexibility of the \ac{led}, i.e., the fixed configuration of the \ac{led} parameters such as the \ac{led}'s \ac{fov} and dimming, prevents the exploitation of the full potential of \ac{vlc} systems. Motivated by its promising features, the \ac{dt} paradigm constitutes an efficient planning, training, and operational tool for enhanced \ac{vlc}-based localization systems. From one perspective, a planning \ac{dt} can be utilized to perform close-to-real testing for optimum \acp{led} placement, in order to realize enhanced coverage and controlled interference, and hence, improved visible light sensing and localization. From another perspective, a training twin can be used as a tool to train multiple network agents, in order to enable them to perform efficient control of the \acp{led} configuration, by manipulating the transmission power and the \ac{fov}, according to different critical cases pertaining to the object different mobility scenarios, blockage by different object sizes, the size of the target, the interference to and from adjacent light diodes, and ambient light interference. Furthermore, the continuous reconfiguration of the \ac{led} parameters should be imperceptible by the human eyes, and hence, agents' training should take into account the uniformity of the illumination as well. It is not hard to tell that such a complex and adaptive scenario requires dedicated computing resources with sophisticated \ac{ml} algorithms, which are difficult to be accurately performed at the physical level alone, thanks to the \ac{dt} for making such a reconfigurable framework feasible, for high resolution localization in visible light environments. This in addition to offering a simple platform for the acquisition of high-quality  \ac{vlc} data, given that the available \ac{vlc} datasets are limited to particular scenarios.    

\textit{\textbf{Summary:}} \Ac{dt} technology is anticipated to offer valuable capabilities for improved high-frequency communication. This can be achieved through utilizing the \ac{ct}, with the multi-modal data acquired from it, to enable optimized system design, where the behavior of high-frequency antennas, transmitters, and receivers, can be tested and accordingly, system parameters pertinent to antenna placement, transmission power, and modulation schemes, can be fine-tuned for maximized performance and coverage. Furthermore, due to their high susceptibility to blockage and interference, multi-modal data acquired from the \ac{dt} can enable improved contextual awareness and allows the identification of active and idle users within the surrounding environment, and therefore, helps in identifying the most effective beamforming design to maximize signal quality and minimize signal loss.

\subsection{Non-terrestrial Wireless Communication}

\subsubsection{Satellite Communications}

\Ac{6g} wireless networks are expected to be designed vertically, incorporating the ground, air, and space layers to provide a ubiquitous coverage with improved reliability and reduced latency. Therefore, \ac{leo} satellite constellations have been deemed as an efficient communication paradigm, and have been identified as an enabler for different use-cases and applications, particularly in smart cities, which require a city-wide on-demand coverage to connect a massive number of users at separated geographical points, while ensuring sustainable operation. A main critical design aspect is the high mobility of \ac{leo} satellites, compared to a ground user, and therefore, frequent handover and service interruption become limiting factors in satellite wireless communications. This is particularly observed in critical applications, e.g., autonomous driving, where service interruption might cause life-threatening incidents. The currently proposed handover schemes for satellite communications rely on optimizing a single metric to enable frequent handover. Such approaches are considered unscalable and don't yield the optimum handover strategy. Rather, multiple attributes need to be considered to ensure optimized handover, and hence, improved \ac{qos} for ground users, opening the door for the \ac{dt} technology to enable enhanced handover strategy design, while considering a wide range of system parameters, without draining the limited vehicles' resources. These parameters include maximum elevation angle, maximum service time, nearest location, handover times, satellite-satellite content delivery, and satellite speed, to name a few. In specific, one can resort to \ac{dt} for performing data generation, computationally-heavy operation, and on-demand operation for multiple \ac{leo} satellites and with considering the ground users' \acp{kpi}. This is due to the capabilities of \ac{dt} to model an entire networked city from the ground to the space, and to ensure the availability of the required synthetic data for the purpose of optimum model configuration, and hence, handover optimization. Recent results demonstrated that the exploitation of \ac{dt} in handover design for satellite communication can potentially reduce the transmission delay between satellite and terrestrial nodes, as well as improving the data delivery quality \cite{zhao2022interlink}.

\subsubsection{\textcolor{black}{Internet-of-Drones}}

\Acp{uav} have attracted a significant attention over the last decade, and have found their applications in wireless networks, such as surveillance, object tracking and coverage extension. This is particularly promising when a swarm of \acp{uav} perform particular tasks in a collaborative manner \cite{lv2021digital}. Hence, in order to implement an efficient \ac{uav}-based network, a swarm of \acp{uav} should be involved, due to the limited capabilities of \acp{uav}. While a swarm of \acp{uav} provides improved coverage and enhanced services, a large number of networked \acp{uav} are needed to cover a relatively large geographical area. This results in an increased management difficulty, and exacerbates the inter- and intra- \ac{uav} coordination complexity. Furthermore, to satisfy the needed reliability and latency requirements, an optimized cross-layer scheduling, that takes into account several \acp{kpi} in addition to the \ac{uav} status, including the instantaneous available resources at each \ac{uav}, as well as the available network resources. Within this context, \ac{dt} technology can play an important role in enabling a live control over \ac{uav}-based delivery systems. Given that all information pertaining to the network and the \acp{uav} for different scenarios will be decoupled at the \ac{ct}, the \ac{dt} can be leveraged to realize cross-layer multi-objective network-wide optimization, in order to efficiently manage the network resources, in terms of spectrum and power allocation, and accordingly, to select the optimum data routing and \ac{uav} trajectory design. Furthermore, the \ac{dt} can develop efficient schemes in order to overcome the single-point of failure in cooperative \acp{uav}, due to energy outage or a \ac{uav} failure. This can be achieved through two ways: 1) multiple agents can be trained over all possible failure scenarios, to perform data re-routing and \ac{uav} handover when a failure is expected to happen, to avoid service interruption and delivery delays. 2) Another way is to employ an operational interactive twin that continuously senses the network and \acp{uav} status, and performs online decision making according to the received information. 

\textit{\textbf{\textcolor{black}{Summary:}}} \textcolor{black}{With the several challenging issues pertaining to satellite operations and data (demonstrating the satellites behavior, or space-based transceivers) acquisition, \ac{dt} can assist in emulating various use-cases to represent several network scenarios, and therefore, generate the needed data sets to identify potential bottlenecks, optimize system parameters, and improve overall system efficiency. By leveraging this synthetic data, \ac{dt} can be used after for studying areas of coverage gaps and link budgets, taking into consideration satellite orbits, antenna patterns, and atmospheric conditions, allowing improved beamforming and constellation design. This can be further exploited for maintenance purposes, where realistic scenarios related to satellite operations, mission planning, and emergency situations can be generated at the \ac{ct}, to develop response strategies, ultimately improving operational readiness and efficiency of satellite networks. }

\subsection{Beyond Massive MIMO}
While massive MIMO technology has demonstrated a successful approach towards enabling improved wireless communication and increased network capacity, through antenna densification at the base-station, it is yet constrained by the maximum number of antennas that can be deployed at each base-station. This constitutes a bottleneck in the network capacity expansion in 6G networks. The vision of smart buildings concept in the era of beyond 5G networks is not limited to implementing smart technologies inside buildings, rather, the vision is to have buildings that are smart from inside and outside, to serve multiple indoor and outdoor applications. The recent technological advancements in smart materials and \ac{ris} have made this vision possible, where large intelligent surfaces can be mounted at the buildings facades in order to perform different functionalities, and support outdoor use-cases. It is worthy to note that the emergence of intelligent surfaces, which comprise a number of reflective elements with reconfigurable amplitude and phase, has revolutionized the way how wireless communications is performed, where an intelligent control over the wireless environment can be achieved through a proper tuning of the \ac{ris} elements. A planning \ac{dt} can be employed to design the optimum \acp{ris} placement and orientation scheme, in a way to serve multiple users that are randomly distributed. In particular, different \ac{ris} locations and orientation can be tested over the \ac{ct}, in terms of coverage, interference control, and security, before the real implementation in the physical environment. This is particularly helpful in large surfaces, which require huge efforts and costs in order to be placed at buildings, and therefore, the selected placement and orientation of the \ac{ris} should be guaranteed to be optimum. From a different perspective, an operational twin can be leveraged in order to realize an on-demand control over the \ac{ris} configuration. It should be highlighted that large intelligent surfaces are capable of serving multiple users and enabling different applications, simultaneously, when their parameters are properly tuned. For example, multi-user support can be achieved, simultaneously with performing smart control over the street lighting system and weather sensing, through proper configuration of the \ac{ris} elements, i.e., which element serves what scenario and what are the optimum amplitude and phase of each element. In order to realize the full potential of \acp{ris} in 6G networks, the latter information cannot be fixed over the network life-cycle, however, it should be adaptive in order to fit the vision of smart networks. Accordingly, the operational twin can handle the heavy computational overhead imposed by the complex optimization and continuous reconfiguration of the \ac{ris} characteristics, in a live manner, and it can provide the required adaptability and flexibility, that are key elements in intelligent systems. 

\textit{\textbf{\textcolor{black}{Summary:}}} 
\textcolor{black}{The main advantage of exploiting \ac{dt} in \ac{ris} applications is the possibility of efficiently design the \ac{ris} network in terms of placement, orientation, and configuration, to accommodate various network and user mobility scenarios. It further facilitates the understanding the multi-\ac{ris} interference through multimodal data, where the latter offer improved contextual awareness and therefore, an enhanced interference detection and mitigation.  Operational \ac{dt} can be also exploited for low-complex and highly reliable real-time configuration of of the \ac{ris} elements, where new network configuration observed at the \ac{ct} can be leveraged to optimize the \ac{ris} design, which can be then reflected at the \ac{pt}.}

\subsection{Wireless caching}

Wireless data caching, where popular data according to users demands are brought closer to edge devices, has shown a tangible improvement in terms of latency, energy, and spectral efficiency. Data caching can follow two approaches, namely, infrastructure-based (where data are cached at the nearest base station) and infrastructure-less (where data are cached at end devices). While the latter enjoys a reduced delay and allows device-to-device communications, the former has higher storage capabilities. In this regard, several mechanisms were proposed to optimize the caching experience at edge users, while taking into consideration different limitations, in terms of cache resources, data popularity profile, users' demands and patterns, and overall network's nodes behaviour. Although some mechanisms have demonstrated robust performance while taking into account one or few of the earlier mentioned requirements, wireless data caching design and optimization is still an open challenge to be addressed in future wireless networks. This is pertinent to the large amount of information that need to be processed in order to enhance the prediction model of data caching. While \ac{ai} algorithms can play an important role in such an optimization problem, the lack of sufficient data sets, that comprise several scenarios of wireless caching and taking into consideration all data caching limitations, limits the accuracy of the developed \ac{ai} models \cite{sheraz2020artificial}. A high-fidelity \ac{dt} can provide a holistic overview of the network caching requirements, and offer a better understanding of the popular data profiles and users needs, and hence, improve the accuracy of the caching prediction. In particular, a high-resolution \ac{ct} can be considered as a rich environment for generating the needed information with the aim to enhance the data caching experience, including geographical, social, and network data, in addition to offering a sophisticated framework for explaining the correlations among different data dimensions, while considering users' historical patterns and demands, and hence, allows improved models training for more accurate data caching design.

\textit{\textbf{\textcolor{black}{Summary:}}} \textcolor{black}{By leveraging real-time synthetic data, \ac{dt} can be exploited as a tool to enable improved content placement, adaptation, and resource management in wireless caching. From the one hand, analyzing real-time data over the \ac{dt}, including cache utilization, network condition, content popularity, and user demands and behavior, efficient resource allocation strategies can be suggested at the \ac{ct}. This can be further exploited for recommending cache updates and eviction, and hence, ensure maximized cache hit rate and reduced latency.}

\subsection{Network Slicing}

The recent progress in network slicing paradigm has come as a consequence of the diverse applications that have emerged and are envisioned to be supported simultaneously by 6G. As its name indicates, network slicing relies on dividing the physical network into multiple separated virtualized segments that share the same physical infrastructure, where each segment (slice) is anticipated to support a particular application. Such a paradigm is the key towards realizing the adaptivity and flexibility required in future wireless generations. Owing to the dynamic nature of wireless networks, in addition to the inherent traffic management and resource allocation challenges, which are pertinent to the virtualization and softwarization in network slicing, \ac{dt} offers a prominent solution for optimized resource allocation in the network slicing paradigm \cite{naeem2021digital}. In particular, multiple \acp{ct} can be designed to represent multiple network slices, and monitor and update the network states using graph theory. It is noted that \ac{gnn} at the \ac{dt} can be leveraged to capture the complex interdependence between multiple slices, which are difficult to quantify in conventional wireless systems due to the irregular topology of graphs, and hence, to obtain the optimum slicing policy that maximizes the network performance \cite{wang2020graph}. Different than relying completely on the physical environment, the \ac{dt} offers a comprehensive replica of the physical environment, where more information can be acquired to enhance the models accuracy, and therefore, strike an optimum end-to-end performance of the network metrics. 

\textit{\textbf{\textcolor{black}{Summary:}}} \textcolor{black}{\Ac{dt} technology offers a powerful approach for efficient network design, resource allocation, performance monitoring, dynamic adaptation, and testing of network slicing configurations and services. In particular, multi-modal data acquired at the \ac{ct} can be leveraged with the real-data from the physical environment to optimize the allocated resources between multiple slices, where the \ac{dt} can be used to identify resource bottlenecks, forecast demand, and dynamically allocate resources to different slices based on their requirements. Furthermore, \ac{dt} can be utilized as a tool to emulate various network conditions, traffic patterns, and user behaviors, in order to verify the efficiency of the network slice configurations before their deployment in the physical network.}

%quantum information processing

%According to today’s physics, the only source of true randomness is quantum. 
%\cite{osti_10359449}

%In the modern viewpoint, quantum computers can be used and trained like neural networks. We can systematically adapt the physical control parameters, such as an electromagnetic field strength or a laser pulse frequency, to solve a problem.

%global-scale quantum internet

%mathematical tools used to model these phenomena are complex (no pun intended) and again rather abstract;

%Classical optical devices lack precision when they operate on single photons. We report a Quantum Digital Twin (QDT) to improve Quantum Key Distribution (QKD) implementations. We show a QDT increasing the Key Exchange Rate under environmental events.

%An important problem connected to entanglement distribution is the handling of the noise on the entangled states. Considering the noisy physical links and other effects of the environment, the received quantum states are noisy. Particularly, the fidelity of the actually created entangled system σ is far from the target fidelity F. To handle the situation, the noisy states should be purified—this process is called entanglement purification. Network optimization is essential in a quantum Internet setting to reduce the purification steps.

\subsection{Integrated Sensing and Communication}

Being two of the main 6G verticals, high performance communication and ultra-reliable sensing were integrated into a unified system, opening up a new paradigm known as \ac{isac}, which allows communication tasks and sensing to share the same hardware and spectrum resources. The latter is performed under the assumption that an acceptable level of performance degradation will be experienced at the two services. Despite its promising potential in achieving on-demand throughput enhancement, improved hardware utilization, and reduced energy consumption, \ac{isac} paradigm has several fundamental constraints that limit its application in current and future wireless networks, in regard to communication reliability and localization accuracy. Driven by the several advantages that can be offered by the \ac{dt}, it is anticipated that \ac{dt} can overcome some limitations experienced in \ac{isac} systems. Motivated by the fact that 6G networks are envisioned to realize a positioning accuracy of sub-centimeters, multi-modal data that are generated by a high-fidelity \ac{dt} can be exploited to acquire more information about objects of interest, and hence, lead to a higher localization and positioning accuracy. Multi-modal localization allows end devices to select the optimum channel and communication methodology given the information obtained from the generated data at the \ac{ct}. From a communication perspective, enhanced context awareness at the \ac{dt} plays an important role in identifying the need for data communication, by predicting the ideal time to communicate information signals. As a consequence, improved network resources utilization and on-demand throughput can be achieved. As a joint advantage for the co-existing systems, the \ac{dt} paradigm can provide high-resolution \ac{csi} acquisition, via vision-aided estimation methods, which highly rely on multi-modal data. High precision \ac{csi} estimation represents a key element toward realizing an efficient integration between communication and localization systems.

\textit{\textbf{\textcolor{black}{Summary:}}} \textcolor{black}{The multimodal nature of data acquired from the \ac{dt} constitutes a promising factor in realizing 3D wireless sensing, in which wireless signals, combined with users’ locations, can be used to reconstruct a 3D image of the wireless environment. Improved wireless sensing can readily contribute to enhancing the communication schemes. In particular, from the reconstructed 3D image, improved beamforming, localization, resource management, etc., can be achieved.}

%\subsection{\textcolor{red}{Mobile Edge Computing/Task Offloading}}

\subsection{Massive Unsourced Random Access}
%classifier which node is active?

As recent wireless technologies are oriented towards enabling smart city applications, with a noticeable increase in the number of IoT devices (in the order of thousands), \ac{mura} paradigm has emerged, where a subset of available devices are assumed to be active simultaneously. While this has come as a result of the need for multi-access support for ultra-high dense networks, it imposes several challenges in terms of \ac{csi} acquisition and the availability of sufficient training sequences, in order to identify the active nodes and their number. As a potential solution, researchers are resorting to exploiting information pertaining to the statistical fluctuation of all nodes' channels to estimate the large-scale component of wireless channels. Nevertheless, extracting such an information requires collecting a large sample-set of received signal power at each node, with the aim to capture the statistical behaviour the random channels, imposing a huge data exchange overhead over resources-limited devices. The unavailability of sufficient data detains the development of accurate \ac{ml} models, in addition to the numerical instabilities experienced in theoretical approaches as the number of devices increases \cite{fengler2019}. Within this context, a high-fidelity \ac{dt} can offer an approach for coherent activity detection in \ac{mura} scenarios, where the historical channel behaviour of a massive number of uncoordinated nodes can be recorded at the \ac{ct}, which can be then leveraged as a seed for \ac{ml} models training. It is worth mentioning that, in \ac{mura}, the system can benefit of an statically-based \ac{ml} model that is built on mathematical models and that rely on statistical behaviors of the training environment. However, such models, e.g., Naive Bayes, requires a large set of high-quality data in order to build the basis for such models. Therefore, over a sufficiently long period of time and considering a wide-range of scenarios and number of total/active nodes, the collected historical data at the \ac{ct} is anticipated to play an important role in facilitating the design of accurate and robust statistical models that can precisely track the statistical behaviour of the unsourced nodes and their channels, and hence, enables the corresponding trained models to operate accurately when implemented in the \ac{pt}. Such an approach significantly reduces the overhead resulted from data exchange for \ac{csi} acquisition in \ac{mura} and hence, offers a low-cost, reliable, and energy-efficient multi-access support in future wireless networks \cite{ruah2022digital}.

\textit{\textbf{\textcolor{black}{Summary:}}} \textcolor{black}{The key element in employing \ac{dt} technology in \ac{mura} scenarios lies in the contextual and situational awareness offered by the \ac{dt}, in which multimodality plays an essential role in capturing active and idle users. This not only allows to identify the number of active users, but also to recognize which users are active, quantify access delay, collision probability, and resource utilization, enable accurate vision-aided channel estimation, and thereby, realize reliable signal detection in \ac{mura} systems. According to the channel conditions of the random users, resource can be allocated in an optimized manner.}

\subsection{Ultra-Dense HetNets}

The recent urban development in terms of intelligent city applications has corroborated the fact that future wireless networks will run over ultra-dense \ac{hetnets}. This has motivated the need for a significant network capacity expansion, in order to meet the demands of future dense smart cities. Ultra-dense \ac{hetnets} are characterized by the extreme densification of base stations, the massive number of heterogeneous nodes with diverse access technologies, and varied cell sizes. While such networks can offer a reasonable capacity enhancement, they raise several concerns pertaining to hardware miniaturization design, intra- and inter-cell coordination, resource allocation, as well as interference management. In this regard, \ac{dt} technology can play a useful role in ultra-dense \ac{hetnets} through leveraging the multimodal information that can be extracted from the \ac{ct} in order to achieve the optimum performance of ultra-dense \ac{hetnets}. In specific, in such networks, small-sized base stations need to be equipped with various sensing, computing, and decision-making capabilities, which, with the massive increase of network sizes and complexity, can potentially drain the network resources. On the other hand, virtual ultra-dense \ac{hetnets} can accurately mimic the behavior of real ones, and hence, with extremely reduced resource consumption, can provide a promising solution to optimize the configuration of \ac{hetnets} at the physical realm. Furthermore, given the highly dynamic and complex nature of such networks, it is envisioned that the \ac{dt} constitutes a rich platform for multi-dimensional information acquisition, with the aim to realize enhanced awareness and cognition in \ac{hetnets}. This can be equivalent to the deployment of all-senses base station, that are capable of capturing multi-modal data, that can be leveraged to train \ac{ai} models for improved inference, with reduced network overhead. Efficiently trained \ac{ai} models are anticipated to realize fully-aware base stations, and hence, to perform complex network-wide optimization, taking into consideration beamforming design, cell-zooming, resource allocation, cells sizes, interference, and base stations transmit power, for various \ac{hetnets} sizes. It is worth highlighting that, such an optimization framework cannot be solved through conventional theoretical nor \ac{ai}-based approaches, where the former lacks the adaptivity, scalability, and tractability, and the latter fails to provide accurate models and inference relying on the limited data collected from the physical environment.

\textit{\textbf{\textcolor{black}{Summary:}}} \textcolor{black}{Optimizing resource allocation strategies in ultra-Dense HetNets is a main motivation behind considering \ac{dt} technology. In particular, \ac{dt} can be utilized for improved real-time dynamic allocation of spectrum resources, adjusting transmit power levels, and performing interference management techniques to maximize network capacity and improve overall spectral efficiency. The overhead of sensing, computing, and decision-making at the small cells can be reduced through leveraging the virtual twin to perform resource optimization and management.}
%\subsection{\textcolor{red}{Security?}}

%\section{Use-Cases}

%\subsection{\textcolor{red}{Localization and positioning}}

%\subsection{\textcolor{red}{Smart Agriculture}}

%\subsection{\textcolor{red}{Healthcare}}

%\subsection{\textcolor{red}{Mobility as a Service}}

%\subsection{\textcolor{red}{Anomaly Detection}}

%\subsection{\textcolor{red}{Underwater wireless networks}}

%\subsection{\textcolor{red}{AR/VR/XR}}

%\subsection{\color{red}Search and Rescue} 
%\subsection{\color{red}Public Safety}
%first responders - best route - disaster prediction 
\section{Wireless Networks for Digital Twin: A Vision}

\subsection{Semantic Communication}

The rise of the semantic communication paradigm is fueled by its promising advantages that promote the development of several technologies, with strict energy, spectrum, and delay requirements. Semantic communications can be efficiently utilized for sensed information communication between the physical and cyber realms \cite{wang2023semantic}. It should be highlighted that, in order to improve the \ac{qoe} at the \ac{dt}, the sensing granularity should be increased, and therefore, imposing a huge data processing, storage, and communication overhead, directly impacting the \ac{qos} performance of the system. This is more pronounced when considering immersive multimedia data. In this regard, the role of semantic-aware communication within the context of \ac{dt} is two-fold. First, with the aim to reduce the communication overhead resulted from multimedia transfer, semantic communication can be utilized to extract the important meanings (semantics) encapsulated in the data to be transmitted, and share it to and from the \ac{ct}, thereby, extremely reducing the amount of data that need to be transferred. Second, as several multimedia streams are anticipated to be exchanged between the physical and cyber environments, and given that there exists a sort of correlation across different data modalities exchanged, semantic communication can further boost the spectral efficiency through extracting the cross-modal semantics and share them with the \ac{dt} servers, further reducing the communication overhead.

\subsection{Joint Source \& Channel Coding}

While the concept of separation has been one of the basic principles in information theory, the emergence of novel paradigms with delay constraints, including the \ac{dt}, has motivated the research in the field of finite block-length coding and communication. In this regard, \ac{jscc} is considered as a promising solution to support ultra-low latency systems. \Ac{jscc} allows the channel decoders to exploit the residuals of the source decoder (stemmed from the fact that for finite-length blocks, the source is considered redundant), and hence, not only reduce the delay, but maintain an acceptable reliability performance \cite{fresia2010joint}. Within the \ac{dt} paradigm, communicating data between the \ac{pt} and \ac{ct} in ultra-reliable and low-latency fashion constitutes a cornerstone in the implementation of an efficient \ac{dt}. Therefore, \ac{jscc} can potentially contribute to the realization of ultra-reliable low-latency communication in \ac{dt}-empowered wireless networks. Although the joint approach might experience lossy signals recovery process at the \ac{dt}, some techniques can be followed to reduce the recovery loss, including Markov model-based source that is jointly decoded with the channel code.

\subsection{Emergent Communications}

\Ac{m2m} communication represents an instrumental paradigm in current and future wireless generations, and due to its machine-centric nature, it offers potential advantages to the \ac{dt}, including enhanced spectral and energy efficiency and reduced delay. A promising tool, emergent communication has been regarded as a key to enable machines to communicate and interact meaningfully, without human intervention, in order to perform particular tasks \cite{lowe2019pitfalls}. These tasks are generally characterized by their high level of coordination requirements, and therefore, emergent communication can offer an intelligent interaction between machines at the \ac{pt} and \ac{dt} servers to perform joint tasks in a coordinated manner. While there might be some delay experienced, once emergent agents reach to an agreement with respect to their communication protocol, several tasks can be achieved with low-latency between the two realms. If efficiently evolved, emergent agents can help with relieving the communication overhead, and save a considerable amount of network resources.

\subsection{Waveform Design}

As elaborated earlier, the \ac{dt} paradigm will impose strict requirements pertinent to data rate, reliability, spectral efficiency, power consumption, and latency. Therefore, it is anticipated that current waveform designs might fail to deliver the needed \ac{qos} for the \ac{dt} paradigm \cite{demir2019waveform}. Being sensitive to synchronization errors, \ac{ofdm} can experience some limitations when implemented within the context of the \ac{dt}, where a huge volume of data is generated from the \ac{dt} operations and communicated, and hence, perfect synchronization is a challenging task to be achieved. Therefore, a joint waveform design of sensing and communication is most likely to define the future of \ac{dt}-centric waveforms, in order to strike a balance between these two essential functionalities in the \ac{dt}. The latter can be designed through considering various approaches, e.g., spatial modulation, joint time/frequency design, and waveform optimization approaches. Such approaches offer enhanced degrees of freedom, and mutual gains between the sensing and communication capabilities. Different than frequency-based waveforms, spatial multiplexing approaches, e.g., \ac{oam}, unfold several advantages to the \ac{dt} with respect to data rates, particularly with the consideration of massive MIMO schemes. This is due to the fact that in order to reap the full potential of \ac{oam}, a massive number of antennas need to be employed. The latter approach of waveform design can incorporate index modulation in order to ensure reliable \ac{pt}-\ac{ct} communication with high data rates and low power consumption \cite{affan2021performance}.

\subsection{Non-coherent Communication Systems}

Over the last few decades, coherent systems, where accurate \ac{csi} can be acquired at the transmitter/receiver for signals recovery purposes, have been the a pillar in all wireless generations. Although non-coherent systems have been extensively discussed in various use-cases, their applicability to the earlier wireless generations was limited due to their degraded performance, compared to their coherent counterpart. Nevertheless, with the emergence of the \ac{dt} paradigm and related technologies that incorporate massive campaigns of data transfer in a live manner, coherent systems represent an obstacle in accommodating the latency and spectral efficiency requirements of \ac{dt}-enabled networks. This is stemmed from the fact that the task of \ac{csi} acquisition for all transmitted sensed information from and to the \ac{ct} is close to impossible, within the network resources constraints. Therefore, there are reignited efforts devoted to explore the potential of blind methods in the \ac{dt} paradigm. For short-length information, e.g., data pertaining to weather, status, etc., maximum likelihood sequence estimation \cite{kailath1969general}, which relies on extracting the correlation between consecutive signals for upcoming signals recovery, is a possible approach. However, as we move to more immersive, larger data, related to multimedia, e.g., 3D videos and images, the performance of such an approach severely degrades. As an alternative, Grassmannian signaling (which was demonstrated to achieve close-to-optimal pilot-free modulation) was deemed as an efficient method for longer information sequences. In Grassmannian modulation, where information is conveyed over tall unitary matrices, which enjoy robust sub-spaces (where each signal is carried) over a wide range of \ac{snr} values.

%\subsection{\textcolor{red}{Communication Protocols}}

\section{Open Research Directions}

\subsection{Trust \& Security}

Being the main controller that manages all network elements, it is very important to ask to what extent we can trust the \ac{dt}. Ensuring a perfectly secured \ac{dt} operation is one of the most critical design aspects that need to be investigated. Specifically, in order to maintain smooth, uninterrupted operations and trusted decisions, the research should put huge efforts in developing secure data communication and authentication schemes that guarantee a trusted twin. This open research issue is of a particular interest in mission-critical applications, due to the severe consequences of security breaches in such scenarios. While blockchain has been proposed as a mean to ensure security at the \ac{dt}, it is yet to be tested whether the heavy operations and high latency in blockchain can be tolerated by the \ac{dt} paradigm. 

\subsection{Data Communication}% with the twin

With the successful deployment of \ac{dt} in wireless networks, it is anticipated to have a huge amount of generated data from distributed sensors to be communicated instantly with the \ac{ct}, imposing challenges on the communication resources between the physical and cyber domains. This necessitates the development of ultra-low latency and high data rate communication paradigms. Although \ac{thz} communications might be seen as a good candidate, it is yet unclear to what degree \ac{thz} systems can provide the necessary coverage for on-demand interactive twin, particularly with high probability of blockage. 

\subsection{To What Extent We Can Go?} 

As a holistic operational twin of a wireless network is still a visionary concept, it is important to lay down a theoretical foundation for exploring the limitations of \ac{dt}-enabled wireless networks from different angles. Specifically, it is essential to revisit the current communication-theoretic models, and to explore whether they can be leveraged to evaluate the performance of \ac{dt}-enabled 6G. Hence, it is necessary to develop a solid, yet scalable and versatile enough mathematical foundation for investigating the capabilities of \ac{dt}-empowered wireless networks, in addition to identifying the system bottlenecks and setting the performance borders at a large-scale. Also, such frameworks assist with further quantifying the benefits that can be reaped, and open new opportunities of innovation, improvement, and development.  

\subsection{Advanced AI Algorithms} 

With \ac{ai} being the engine for activating \ac{dt} and enabling it to be an efficient orchestrator for the network, it is of paramount importance to ensure that the employed \ac{ai} algorithms fit well within the needs of \ac{dt}-enabled 6G, in terms of latency, accuracy, and energy efficiency. It is yet to be revealed whether existing \ac{ai} algorithms will be able to capture the inherent relationships of the physical dynamics and their impact on the wireless environment, and therefore, to develop an understanding of all network functionalities, variations, and interactions over different time periods. With the strong belief that the \ac{dt} will constitute the brain of the network in the future, it is essential to ensure that the trained \ac{ai} models will enable the \ac{dt} to have cognition capabilities, including reasoning and inference. %This opens new horizons for exploring sophisticated \ac{ai} algorithms and models, that will serve the needs of the \ac{dt}. 

\subsection{Quantum Digital Twin}

The advent of quantum computing, that is characterized by high-dimensional quantum states (a.k.a Qubits), has revolutionized the way how classical computers operate, introducing a new era of computing capabilities and functionalities. In the wireless communication domain, the emergence of the quantum domain offers a way to break the capacity limits set by conventional wireless systems, that rely on the basic 0/1 state space \cite{cozzolino2019high}. In addition to the improved channel capacity, quantum communication has a unique robustness to noise and high immunity to eavesdropping. While the research in quantum communications has witnessed a noticeable progression, challenges pertaining to quantum teleportation, quantum error correcting codes, entanglement purification, and the quantum repeater, are yet to be further explored in order to enable long-distance quantum communication. The fusion of \ac{dt} and quantum computing paves the way for several advantages that can be reaped from the two schemes, where the interplay of the classical \ac{dt} paradigm with the quantum computing and processing capabilities (including quantum machine learning) promising to offer real-time simulation, monitoring, and control of highly-complex, highly-dynamic, and interconnected elements, with the aim to achieve the ultimate vision of quantum internet.  

\subsection{Immersive Communication}

The prevalence of virtual and augmented reality devices, in addition to the availability of computing resources and high-resolution equipment, have given a rise to the immersive communication paradigm, which relies on exchanging natural haptic signals with remote devices. The key element in immersive communication is the ability of wireless nodes to interact with remote environments, and to detect and quantify this interaction through exploiting all-sense features. In other words, haptic perception of participating nodes in remote environments should enjoy high resolution to achieve the required QoE. In order to realize the ultimate immersion, several KPIs are imposed on future wireless networks, in terms of communication latency (sub-millisecond) and throughput, where the latter is characterized by the resolution quality, color depth, and frame-rate. A super-resolution \ac{dt} is envisioned to be the key towards realizing fully immersive communication, through offering a holistic framework for multi-sensory data acquisition, with reduced latency and improved resource management. %The research on \ac{dt}-enabled immersive communications is still scarce, and worthy to explore, in order to realize the all-sense vision of future wireless networks.

\subsection{Synchronization vs. Delay vs. Accuracy}

The development of a high-fidelity \ac{dt} is constrained by three main factors, namely, shared data granularity to ensure perfect agreement between the physical and cyber twins, perfect synchronization, and the inference and update latency between the two realms. While the three factors are essential, and highly affect the network performance at the physical twin, each contribute to the successful realization of particular twin modes. For example, low latency can be tolerated in some scenarios that don't require live inference, as long as the two twins are in a perfect sync, meaning that the changes experienced at the \ac{pt} need to updated at the \ac{ct} in a synchronized fashion, in order to perform long-term network configuration. On the other hand, ultra-low latency is necessary in operational twins, where decisions made at the \ac{ct} need to be reflected to the \ac{pt} in a live manner. Yet, a high-fidelity \ac{dt} requires continuous data sharing between the two twins, demanding high throughput and yielding increased delay. %Such a trade-off problem demands the development of novel communication paradigms that are capable of striking a balance between the three different requirements. 

\section{Conclusion}
\label{sec:conclusion}

In this article, we laid down the foundation for integrating wireless technologies with \ac{dt} paradigm, and overviewed the technological trends that have manifested themselves as key enablers for \ac{dt}-assisted 6G. We further outlined the opportunities offered by the \ac{dt} in enhancing the performance of wireless networks, and we shed light on how existing communication systems need to be revisited in order to guarantee that future wireless networks will be capable of supporting the needs of the \ac{dt}. Finally, we highlighted several potential research directions for further exploration of this disruptive innovation.  

\bibliographystyle{IEEEtran}
\bibliography{Refs}

\section*{Biographies}
\textbf{Lina Bariah} (lina.bariah@ieee.org) is a Senior Researcher at the Technology Innovation Institute, Abu Dhabi. %She is an IEEE Senior Member. She serves as an Associate Editor for the IEEE Communication Letters, and the IEEE Open Journal of the Communications Society.  

\textbf{Hikmet Sari} (hsari@ieee.org) is a Professor with the Nanjing University of Posts and Telecommunications (NJUPT). %He is an IEEE Life Fellow. He was a recipient of the Andre Blondel Medal, the Edwin H. Armstrong Achievement Award, the Harold Sobol Award.

\textbf{M{\'e}rouane Debbah} (Merouane.Debbah@tii.ae) is a Professor at Khalifa university, Abu Dhabi. %He is an IEEE Fellow, a WWRF Fellow, a Eurasip Fellow, an AAIA Fellow, an Institut Louis Bachelier Fellow, and a Membre émérite SEE. He has received more than 20 best paper awards.

\end{document}